\documentclass[preprint2]{aastex}
\usepackage{epsf,amsmath}
\usepackage{amssymb}
\usepackage{array,tabularx}

\begin{document}

\title{Luminosity function of binary X-ray sources calculated using the Scenario Machine}

\author{A.I. Bogomazov}

\affil{Sternberg astronomical institute, 119992, Universitetskij
prospect, 13, Moscow, Russia} \email{a78b@yandex.ru}

\author{V.M. Lipunov}

\affil{Sternberg astronomical institute, 119992, Universitetskij
prospect, 13, Moscow, Russia} \email{lipunov@xray.sai.msu.ru}

\begin{abstract}
Using the ``Scenario Machine'' we have carried out a population
synthesis of X-ray binaries for the purpose of modelling of X-ray
luminosity functions (XLFs) in different types of galaxies: star
burst, spiral, and elliptical. This computer code allows to
calculate, by using Monte Carlo simulations, the evolution of a
large ensemble of binary systems, with proper accounting for the
spin evolution of magnetized neutron stars.

We show that the XLF has no universal type. It depends on the star
formation rate in the galaxy. Also it is of importance to take
into account the evolution of binary systems and life times of
X-ray stages in theoretical models of such functions. We have
calculated cumulative and differential XLFs for the galaxy with
the constant star formation rate. Also we have calculated
cumulative luminosity functions for different intervals of time
after the star formation burst in the galaxy and curves depicting
the evolution of the X-ray luminosity after the star formation
burst in the galaxy.
\end{abstract}

\keywords{binaries: close --- binaries: general --- X-rays:
binaries
--- X-rays: general}

\section{INTRODUCTION}

The evolution of the X-ray luminosity of galaxies was predicted by
\citet{tatarintzeva1989a}\footnote{See also \citet{lipunov1996b}}.
They studied X-rays come from X-ray binary stars. The evolution of
the X-ray luminosity $L_\delta(t)$ was calculated assuming the
$\delta$-function shape for the star formation rate (simultaneous
birth of the stars). The evolution of the X-ray luminosity of the
galaxy with an arbitrary star formation rate $\phi(t)$ can be
represented as

\begin{equation}
 L(t)=\int^{+\infty}_{-\infty} L_\delta(t-\tau)
 \phi(\tau)\textit{d}\tau\label{form1},
\end{equation}

The evolution of the total X-ray luminosity after $t>2\times 10^9$
years (long time scale) from the star formation burst can be well
fitted by power law \citep{tatarintzeva1989a}$^1$:

\begin{equation}
\label{tatar} L(t) \approx 3 \cdot 10^{40}
\left(\frac{N}{10^{12}}\right)\left(\frac{t}{10^9\mbox{yr}}\right)^{-1.56}\mbox{erg}\cdot\mbox{s}^{-1},
\end{equation}

\noindent here $N$ is the total number of the stars in the galaxy.

\citet{lipunov1996a} studied the evolution of stellar populations
after the star formation burst occurring in the conditions similar
to the Milky Way, in the central part of the galaxy, on a
timescale of 10 Myr. Their results include a number of X-ray
transients (each consisting of a neutron star and a main sequence
star), super accreting black holes, and binaries consisting of a
black hole and a supergiant, as functions of time. They showed
that absolute and relative numbers of massive binary systems
including neutron stars and black holes can serve as a good
indicator of the age of the star formation burst.
\citet{popov1998a} also made fits to dependencies $N(t)$ for
different types of objects, where $N(t)$ is the number of sources,
$t$ is the time after the star formation burst.

\citet{vanbever2000} combined their close binary population number
synthesis code with the formation mechanism of X-radiation in
young supernova remnants and in high mass X-ray binaries. They
demonstrated that the impact of interacting binaries is
substantial.

% \citet{sipior} presented an X-ray binary population synthesis
% model. They uses it to simulate the evolution of X-ray binaries
% formed in a burst of star formation of duration 20 Myr and
% star-formation rate $10 M_{\odot}$ yr$^{-1}$. The X-ray luminous
% phase in their model persists long after the star-formation
% episode has ended. This result is insensitive to the poorly
% constrained values of the initial mass function and the average
% mass ratio between accreting and donor stars.

Numerous point-like extragalactic X-ray sources were discovered
during last years due to {\it Chandra} (see e.g.
\citet{muno2004a}, \citet{grindlay2005a}) and {\it XMM -Newton}
(see e.g. \citet{kong2003a}, \citet{georgakakis2004a},
\citet{georgantopoulos2005a}) missions. Some authors
\citep{grimm2002a,grimm2003a,gilfanov2004a,kim2004a} report about
power law X-ray luminosity function:

\begin{equation}
\frac{dN}{dL} \sim L^{- \alpha} \times SFR, \alpha \approx 1.5,
\label{dn}
\end{equation}

\noindent where $SFR$ is the star formation rate.

These data were discussed by \citet{postnov2003a} from theoretical
point of view.

\citet{grimm2003a} realized that, within the accuracy of the
presently available data, a linear relation between high mass
X-ray binaries (HMXB) number and star formation rate ($SFR$)
exists. They suggest that the relation between $SFR$ and
collective luminosity of HMXBs is non-linear in the low-$SFR$
regime, $L_x\sim SFR^{\sim 1.7}$, and becomes linear only for a
sufficiently high star formation rate, $SFR \gtrsim 4.5 M_{\odot}$
yr$^{-1}$ (for $M>8M_{\odot}$). Also they obtained the universal
luminosity function of HMXBs and fitted the combined luminosity
function of M82, Antennae, NGC 4579, 4736 and Circinus using a
maximum-likelihood method with a power law with a cut-off at
$L_c=2.1\cdot 10^{40}$ erg s$^{-1}$ and normalized the result to
the combined $SFR$ of the galaxies. Their best-fitting luminosity
function in the differential form is given by

\begin{equation}
\frac{dN}{dL_{38}}=(3.3_{-0.8}^{+1.1}) SFR \times L_{38}^{-1.61\pm
0.12} , L<L_c, \label{dn2}
\end{equation}

\noindent where $L_{38}=L/10^{38}$ erg s$^{-1}$ and $SFR$ is
measured in units of $M_{\odot}$ per year.

\citet{zezas2004a} presented the X-ray luminosity function of the
Antennae galaxies based on 8 observation performed with {\it
Chandra}, 7 of them were obtained between January 2001 and
November 2002. After combining all observations they detect a
total of 120 sources down to a limiting luminosity of $\sim 2\cdot
10^{37}$ erg s$^{-1}$. Authors suggested that comparison between
the XLFs of the individual observations showed that they are not
affected by source variability. The cumulative XLF of the coadded
observations was represented by a single power law $N(>L)\sim
L^{-0.52^{+0.08}_{-0.33}}$. There was an indication for a 'bump'
at $\sim 10^{38}$ erg s$^{-1}$, but at this point its significance
was not clear. If this bump is statistically significant it could
be evidence for Eddington limited accretion on compact objects or
anisotropic emission from the accretion disk \citep{zezas2002a}.

\citet{belc2004} constructed synthetic X-ray binary populations
for direct comparison with the X-ray luminosity function of NGC
1569 observed with {\it Chandra}. They produced hybrid models
meant to represent the two stellar populations: one old and
metal-poor, with continuous star formation for $\sim 1.5$ Gyr; and
another a recent and metal-rich population. They found that for
typical binary evolution parameters, it is possible to quite
closely match the observed XLF shape.

Our critical points concerning both observational and theoretical
aspects is in that that there is no observed universal luminosity
function because:

\begin{enumerate}

\item Number of bright X-ray binaries is very small per
 galaxy.

\item We do not know real X-ray luminosity due to high variability
of binary X-ray sources, on scales from seconds up to 100 years.

\end{enumerate}

There is no simple (with one slope) theoretical universal
luminosity function because:

\begin{enumerate}

\item X-ray population is the mix of different types of binaries
with different mass exchange types.

\item Number of the systems with definite luminosity depends on
spin evolution of a neutron star which has no direct connection to
mass of its companion.

\item Theoretical arguments for universal function being at
present time are not quite correct, because they exclude
life-times (which depend on optical companion mass) of binary
stars in accretion stage \citep{postnov2003a}.

\end{enumerate}

We stress that it is of great importance to take the spin
evolution of NSs into account. Quantity of accreting neutron stars
which give their contribution to the luminosity function is
determined by their magnetic fields and spin periods. Neutron
stars can be in a non-accreting state (propeller, ejector, see for
details \citet{lipunov1992a}). This circumstance usually is not
taken into account in population synthesis models.

We must observe much more sources and determine their types to
make correct luminosity function. In any case XLFs must have
different slope for different types, ages and star formation
histories in galaxies.

Ultra luminous X-ray sources (ULXs) with $L_x > 10^{39} \mbox{erg
s$^{-1}$}$ have been discovered in great amounts in external
galaxies with {\it ROSAT}, {\it Chandra} and {\it XMM-Newton}.
\citet{rappaport2005a} carried out a theoretical study to test
whether a large fraction of the ULXs, especially those in galaxies
with recent star formation activity, can be explained with binary
systems containing stellar-mass BHs. To this end, they have
applied a unique set of binary evolution models for BH X-ray
binaries, coupled to a binary population synthesis code, to model
the ULXs observed in external galaxies. They find that for donor
stars with initial masses $\gtrsim 10 M_\odot$ the mass transfer
driven by the normal nuclear evolution of the donor star is
sufficient to potentially power most ULXs. This is the case during
core hydrogen burning and, to an even more pronounced degree,
while the donor star ascends the giant branch, although the latter
phases last only 5 per cent of the main-sequence phase. They show
that with only a modest violation of the Eddington limit, e.g. a
factor of 10, both the numbers and properties of the majority of
the ULXs can be reproduced. One of their conclusions is that if
stellar-mass BH binaries account for a significant fraction of
ULXs in star-forming galaxies, then the rate of formation of such
systems is $3 \cdot 10^{-7}$ yr$^{-1}$ normalized to a
core-collapse supernova rate of $0.01$ yr$^{-1}$.

\citet{king2001} investigated models for the class of
ultraluminous non-nuclear X-ray sources (ULXs) seen in a number of
galaxies and probably associated with star-forming regions. The
assumption of mild X-ray beaming suggests instead that ULXs may
represent a short-lived but extremely common stage in the
evolution of a wide class of X-ray binaries. The best candidate
for this is the phase of thermal-timescale mass transfer that is
inevitable in many intermediate- and high-mass X-ray binaries.
This in turn suggests a link with the Galactic microquasars. The
short lifetimes of high-mass X-ray binaries would explain the
association of ULXs with episodes of star formation. These
considerations still allow the possibility that individual ULXs
may contain extremely massive black holes.

We also would like to remember the old consideration of the
supercritical non-spherical accretion onto magnetized neutron
stars \citep{lipunov1982a,lipunov1982b}. In this case matter falls
to the magnetic poles of the neutron star. Maximum energy release
proves to be $L=46 L_{Edd}(\mu_{30})^{4/9}$, where $\mu_{30}$ --
is the magnetic dipole moment of the neutron star in
$10^{30}$~G~cm$^{3}$.

\section{DESCRIPTION OF MODELS}

\label{sec2}

\subsection{Binaries under consideration and mechanisms of mass accretion}

\label{sec2-sub1}

The ``Scenario Machine'' code includes the next types of the mass
accretion by the compact star:

\begin{enumerate}

\item Accretion from the stellar wind.

\item Accretion from the disk-shaped stellar wind of Be-stars.

\item Mass transfer through the inner Lagrange point during Roche
lobe overflow stage:

\begin{enumerate}

\item On thermal timescale.

\item On nuclear timescale.

\item On magnetic stellar wind timescale.

\item On gravitational radiation timescale.

\end{enumerate}

\end{enumerate}

Induced stellar wind is not included into the program.

Most of the X-ray pulsars in the Milky Way belong to binaries
which consist of the Be-star and the neutron star
\citep{liu2000a,liu2001a,raguzova2005a}. The mass loss by the
Be-star is determined by its rotation. Its mass influences onto
its wind to a lesser degree. At the same time we see a little part
of the X-ray binaries consisting of Be- and neutron stars due to
variability of mass transfer processes and the transient character
of accretion in such systems \citep{heuvel1994a}.

% «десь €вно надо что-то добавить.

So, we should study as much types of X-ray binaries as possible.
This is the list of the systems under our consideration:

\begin{enumerate}

\item NA+I: the accreting neutron star with the main sequence
companion.

\item NA+II: the accreting neutron star with the super giant
companion.

\item NA+III: the accreting neutron star with  the companion
filling its Roche lobe.

\item NA+Be: the accreting neutron star with the Be-star
companion.

\item BH+II: the black hole with the super giant companion.

\item BH+III: the black hole with the companion filling its Roche
lobe.

\item SNA+III: the super accreting neutron star with the companion
filling its Roche lobe.

\item SBH+III: the super accreting black hole with the companion
filling its Roche lobe.

\end{enumerate}

The last two types of systems are taken into consideration for the
purpose of modelling of ULXs. Radiation of such objects can be
strongly collimated (see e.g. \citet{cherepashchuk2005a}) to a
degree $\sim 1^{\circ}$. Also we take into account possibility of
mild beaming (see e.g. \citet{king2001}). If the radiation of the
source is collimated, then we should reduce calculated number of
binaries using formula

\begin{equation}
\label{nc} N_o=\frac{\Omega}{4\pi}N_{c},
\end{equation}

\noindent because we can not see the object if its beam is
directed away from us. We recalculate X-ray luminosity of such
systems using formula

\begin{equation}
\label{lc} L_o=\frac{4\pi}{\Omega}L_{c},
\end{equation}

\noindent in order to obtain the luminosity under the formal
assumption of spherically symmetric radiation. In these equations
$\Omega$ is the doubled solid collimation angle of the radiation,
$L_c$ is the calculated luminosity of the source and $N_c$ is the
calculated number of sources, $L_o$ and $N_o$ are the same
observable values.

We have to say some words about Wolf-Rayet (WR) stars with black
holes or neutron stars. Number of binaries consisting of the
accreting black hole and the WR-star is very small, because
WR-stars have very high velocity of the wind. That is why
accretion disks are not able to form in wide pairs (with orbital
period $\gtrsim 10$ hours; orbital period of Cyg X-3, for example,
is $\approx 5$ hours; see for detailed description
\citet{karpov}). There are no binaries consisting of WR-stars and
accreting NSs, because NSs accelerate their rotation during second
mass exchange (recycling) and therefore become propellers or
ejectors \citep{lipunov1982c} in such kind of binaries.

Note that our conclusions concerning accreting neutron stars with
Be-stars, super accreting neutron stars with non-degenerate stars
filling their Roche lobes, super accreting black holes with
non-degenerate stars filling their Roche lobes have approximate
character, because it is impossible to depict correctly temporal
and angular dependencies of their radiation at present time. Our
calculations show that real luminosity function is compound.

\subsection{List of main evolutionary parameters}

Since the algorithms used in the ``Scenario Machine'' have been
described many times, we shall only note the most important
evolutionary parameters influencing the results of the numerical
modeling of the binaries under consideration. A detailed
description of the ``Scenario Machine'' can be found in the next
works: \citet{lipunov1996b,lipunov1996c,lipunov2007a}.

The initial masses of primary components were varied between
$10M_{\odot}$ and $120M_{\odot}$. To describe also a kind of ULX
objects consisting of a black hole with mass higher than $\sim 100
M_{\odot}$ and an optical star in Roche lobe overflow stage we
have conducted a population synthesis also with lower and upper
limits equal to $120 M_{\odot}$ and $1000 M_{\odot}$
correspondingly.

We assume zero initial eccentricity, all initial mass ratios have
equal probability, initial mass of the secondary star is in the
range $0.1 M_{\odot}$ -- mass of the primary star.

Mass loss by optical stars in the course of their evolution
remains incompletely explored. Despite the fact that is has been
possible to substantially reduce the uncertainties (see, e.g.,
\citep{bogomazov2005a}), no clear justification for a choice of a
standard scenario has emerged. Therefore, we carried out our
computations for two scenarios for mass loss by non-degenerate
stars, which we call A and C. A detailed description of these
models can be found in \citep{lipunov2007a}. Scenario A has a weak
stellar wind. The stellar wind of massive stars (with masses
higher than $15 M_{\odot}$) is higher in scenario C, for
lower-mass stars, scenarios A and C are equivalent. The total mass
loss in any evolutionary stage is higher in scenario C than in
scenario A.

Common envelope stage efficiency $\alpha_{CE}$ is equal to 0.5.

Minimal initial mass of the star which produces a black hole as
the result of its evolution is $25 M_{\odot}$. We assume the
parameter $k_{bh}=M_{bh}/M_{PreSN}$ to be equal to 0.5 (see
\citet{bogomazov2005a} for detailes), $M_{PreSN}$ is the mass of
the pre-supernova star which produced the black hole with mass
$M_{bh}$.

Initial mass of the new-born neutron star is randomly distributed
in the range 1.25 -- 1.44$M_{\odot}$. Maximum mass of the NS
(Oppenheimer-Volkoff limit) equals to $M_{OV}=2.0 M_{\odot}$ in
our calculations. Initial value of the magnetic field of NSs is
assumed to be equal to $10^{12}$ Gs, the field decay time is
assumed to be equal to $10^8$ years. Characteristic kick velocity
of the new-born neutron star we accept to be equal to 80 km
s$^{-1}$ in this work.

We use two different values of collimation angle for supercritical
regimes of accretion: $\alpha=1^{\circ}$ and $\alpha=10^{\circ}$.

\subsection{Result normalization}

Birth frequencies of binaries were calculated using the next
formula:

\begin{equation}
\nu_{gal}=\frac{N_{calc}}{N_{tr}}\frac{1}{M^{1.35}_1},
\label{frsp}
\end{equation}

\noindent here $\nu_{gal}$ is the frequency of birth of a specific
binary system type in a spiral galaxy, $N_{calc}$ is the number of
the systems under our consideration appeared during calculations,
$N_{tr}$ is the total number of binaries which evolution was
calculated, $M_1$ is the minimal initial mass of a star in our
calculations. We treat a spiral galaxy in this case as a galaxy
with constant star formation rate which is defined by the Salpeter
function.

Quantities of the systems in a spiral galaxy were calculated using
equation (\ref{sp}).

\begin{equation}
N_{gal}=\frac{\sum t_i}{N_{tr}}\frac{1}{M^{1.35}_1},\label{sp}
\end{equation}

\noindent here $N_{gal}$ is the quantity of a specific binary
system type in a spiral galaxy, $t_i$ is the life time of the
binary system under consideration.

Along with modeling population in the spiral galaxy we also made
some estimations of evolution of X-ray luminosity function and
total X-ray luminosity in an elliptical galaxy. Quantities of the
systems in the elliptical galaxy were calculated using equation
(\ref{ell}).

\begin{equation}
N_{gal}=N_{calc}\frac{M_{gal}}{M_{ScM}} \left(
\frac{M_{1ScM}}{M_{1gal}} \right)^{-1.35}\frac{\sum t_i}{\Delta
T}, \label{ell}
\end{equation}

\noindent here $M_{gal}=10^{11} M_{\odot}$ is the mass of typical
galaxy, $M_{1ScM}$ is the minimal initial mass of a star in our
calculations, $M_{1gal}=0.1 M_{\odot}$ is the minimal initial mass
of a star, $t_i$ is the life time of a binary system under our
consideration in the range of the time interval between $T$ and
$T+\Delta T$. We treat an elliptical galaxy in this work as a kind
of object in which all stars born at the same time and then evolve
($\delta$-function star formation rate).

\subsection{Constraints on key parameters
of the evolutionary scenario}

Previous estimates of the ranges of parameters determining the
evolution of binaries were obtained by
\citet{lipunov1996c,lipunov1997a}.

Since that time, some new results related to the evolution of
binaries have been obtained, and we carried out additional
computations of constraints that can be applied to the parameters
of the evolutionary scenario. The latest observational estimates
of the kick velocities received by NSs in supernovae explosions
are given by \citet{hobbs2005}, where it is concluded that the
typical kick magnitude is $\sigma=265$ km s$^{-1}$.

An attempt to obtain a more accurate estimate of the mass-loss
efficiency in the common-envelope stage was made by
\citet{dewi2000}, who tried to take into account the concentration
of the stellar material toward the center: $\frac{GM_d(M_d -
M_c)}{R_d\lambda}$. However, they assumed that the efficiency in
the common-envelope stage was $\mu_{CE}=1$. In general, this
parameter is not known accurately. Our coefficient $\alpha_{CE}$
is the product of $\mu_{CE}$ and the parameter $\lambda$
considered by \citet{dewi2000}, which describes the concentration
of the stellar matter toward the center. For this reason, we use
the value of $\alpha_{CE}$ suggested by \citet{lipunov1996c}.

We would like to note one more important circumstance. Ill-defined
parameters of the evolutionary scenario, such as $v_0$,
$\alpha_{CE}$, the stellar wind efficiency, and so on, are
internal parameters of the population synthesis. In the future,
they may be defined more precisely, or their physical meaning may
change: the kick-velocity distribution may turn out not to be
Maxwellian, it may be that the complex hydrodynamics of common
envelopes cannot be described using the parameters $\alpha_{ce}$
and $\lambda$, the mass ratio distribution $f(q)$ may be not a
power law. There exists only one way to verify our results:
comparison of our model predictions with observational data.

For this reason, we suggest two quantities to be compared to test
the model: the ratio of the calculated and observed numbers of Cyg
X-3 systems, and the ratio of the number of binary radio pulsars
with NS companions and the total number of radio pulsars (both
single and binary), $\frac{N_{Psr+NS}}{N_{Psr}}$, where
$N_{Psr+NS}$ is the number of radio pulsar in binary systems with
a neutron star, $N_{Psr}$ is the total number of radio pulsars,
binary and single. To avoid the need to differentiate between
young pulsars and old pulsars that have been accelerated by
accretion, we consider only young pulsars. Note that the observed
value of this ratio is $\sim 0.001$ \citep{atnf}: among more than
1500 known single and binary radio pulsars, only two young pulsars
in pairs with NSs have been discovered (J2305+4707 and
J0737-3039). As a model of a Cyg X-3 type system, we take a binary
containing a BH with WR companion that is more massive than $>7
M_{\odot}$ and having an orbital period 10 hours.

Figure \ref{dpsr} of the calculated number of binaries with a NS
and radio pulsar $N_{Psr+NS}$ and the calculated sum of the
numbers of single and binary radio pulsars $N_{psr}$ depends on
the kick velocity $v_0$. The width of the shaded region reflects
the variation of the efficiency of the common envelope stage
$\alpha_{CE}$ in the range 0.2Ц1.0.

Figure \ref{occo} shows the OCCO criterion \citep{lipunov1996b}
for the ratio $\frac{N_{Psr+NS}}{N_{Psr}}$. The typical kick
velocity $v_0$ is plotted along the horizontal axis. The width of
the shaded region reflects the variation of the efficiency of the
common envelope stage $\alpha_{CE}$ in the range 0.2Ц1.0. The
observed value of $\frac{N_{Psr+NS}}{N_{Psr}}$ is $\sim 0.001$.

As seen from Figs. \ref{dpsr} and \ref{occo}, the characteristic
value of kick velocity $v_0$ cannot exceed $\approx 200$ km
s$^{-1}$. By this reason we make use of the results of paper
\citep{lipunov1997a}.

\begin{figure}
\plotone{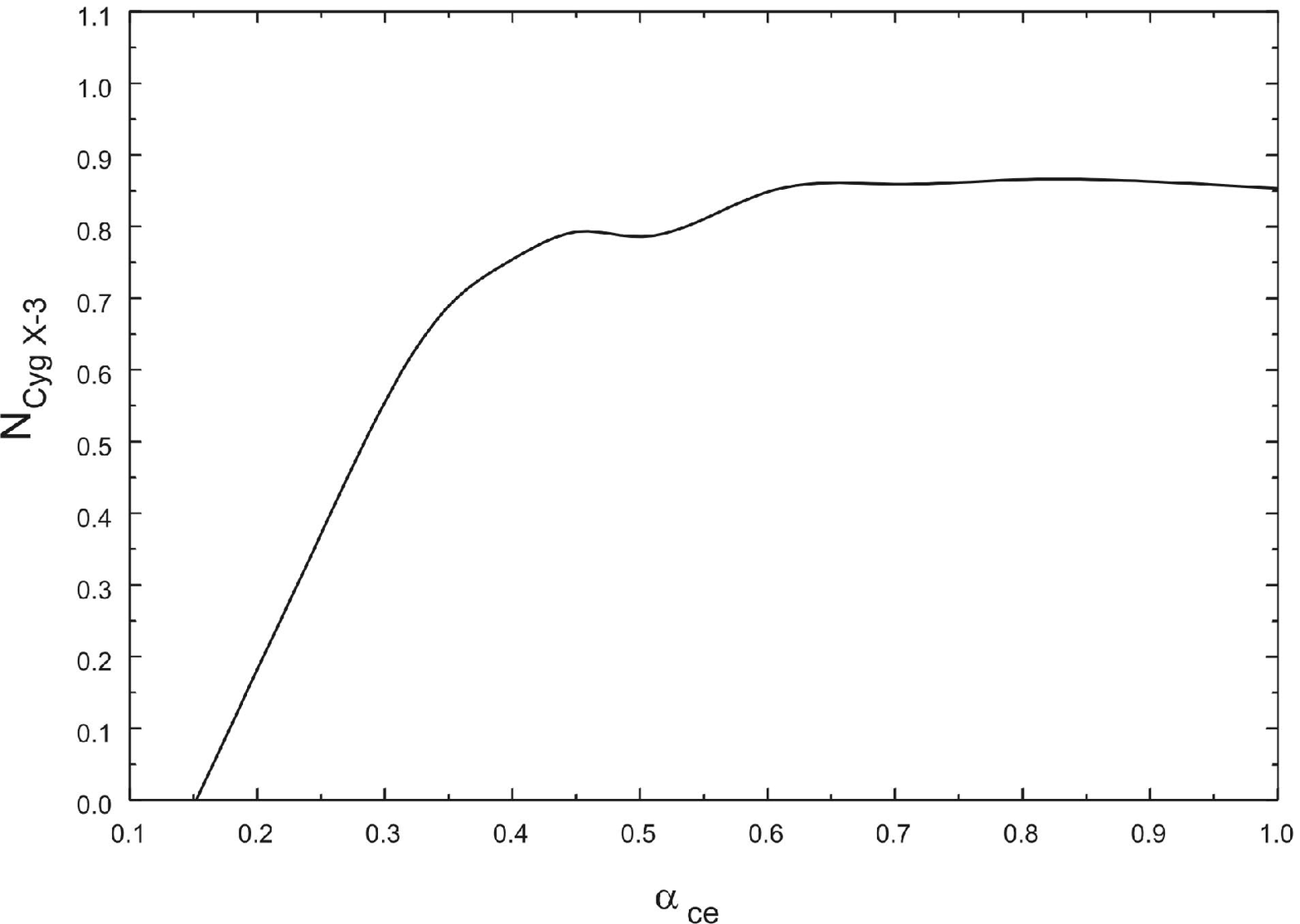}\figcaption{ Calculated number of Cyg X-3 type
systems in the Galaxy as the function of the common envelope stage
efficiency $\alpha_{CE}$. }\label{cygx3}
\end{figure}

\begin{figure}
\plotone{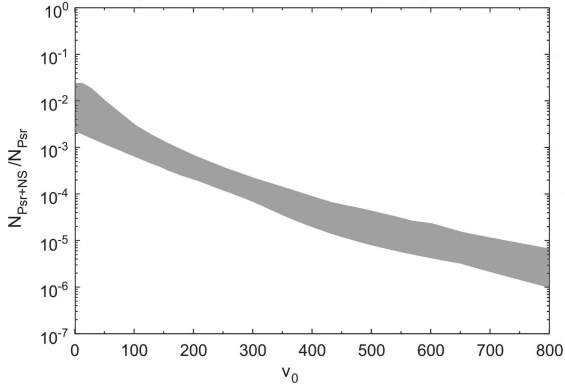}\figcaption{ This figure shows how the ratio
$\frac{N_{Psr+NS}}{N_{Psr}}$ depends on the kick velocity $v_0$.
Here $N_{Psr+NS}$ is the calculated number of binary neutron stars
with radio pulsars and $N_{Psr}$ is the calculated number of all
radio pulsars, binary and single. ``Width'' of the filled area
depicts various values of $\alpha_{CE}$ in the range $0.2-1.0$.
}\label{dpsr}
\end{figure}

\begin{figure}
\plotone{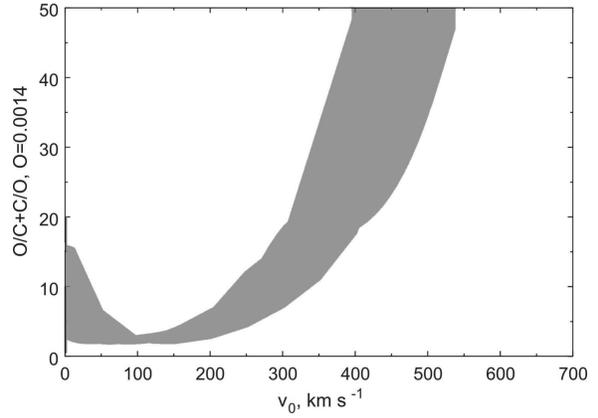}\figcaption{ This figure shows OCCO criterion
\citep{lipunov1996b} for the ratio $\frac{N_{Psr+NS}}{N_{Psr}}$,
$v_0$ is the characteristic kick velocity. ``Width'' of the filled
area depicts various values of $\alpha_{CE}$ in the range between
0.2 and 1.0. Observational value of the ratio
$\frac{N_{Psr+NS}}{N_{Psr}}$ is $\sim 0.001$. }\label{occo}
\end{figure}

Figure \ref{cygx3} shows the number of Galactic Cyg X-3 systems in
our model as a function of the common envelope efficiency. This
figure shows that we can essentially exclude values
$\alpha_{CE}<0.3$.

\section{RESULTS AND CONCLUSIONS}

Four simulations runs were performed, each simulating the
evolution of $1\cdot 10^7$ binary systems. Two of them were
performed with weak stellar wind (stellar wind type A), and other
models with reasonably high stellar wind (stellar wind type C). In
each of these cases we made our calculations using two value areas
of initial mass of the primary star in Salpeter's power law: in
the range $10-120M_{\odot}$ for investigations of all types of
systems under consideration, and in the range $120-1000M_{\odot}$
to qualitatively depict only ultra luminous objects consisting of
super accreting intermediate mass black holes with companions
filling their Roche lobes.

\begin{figure*}
\epsfxsize=1.0 \textwidth\centering\epsfbox{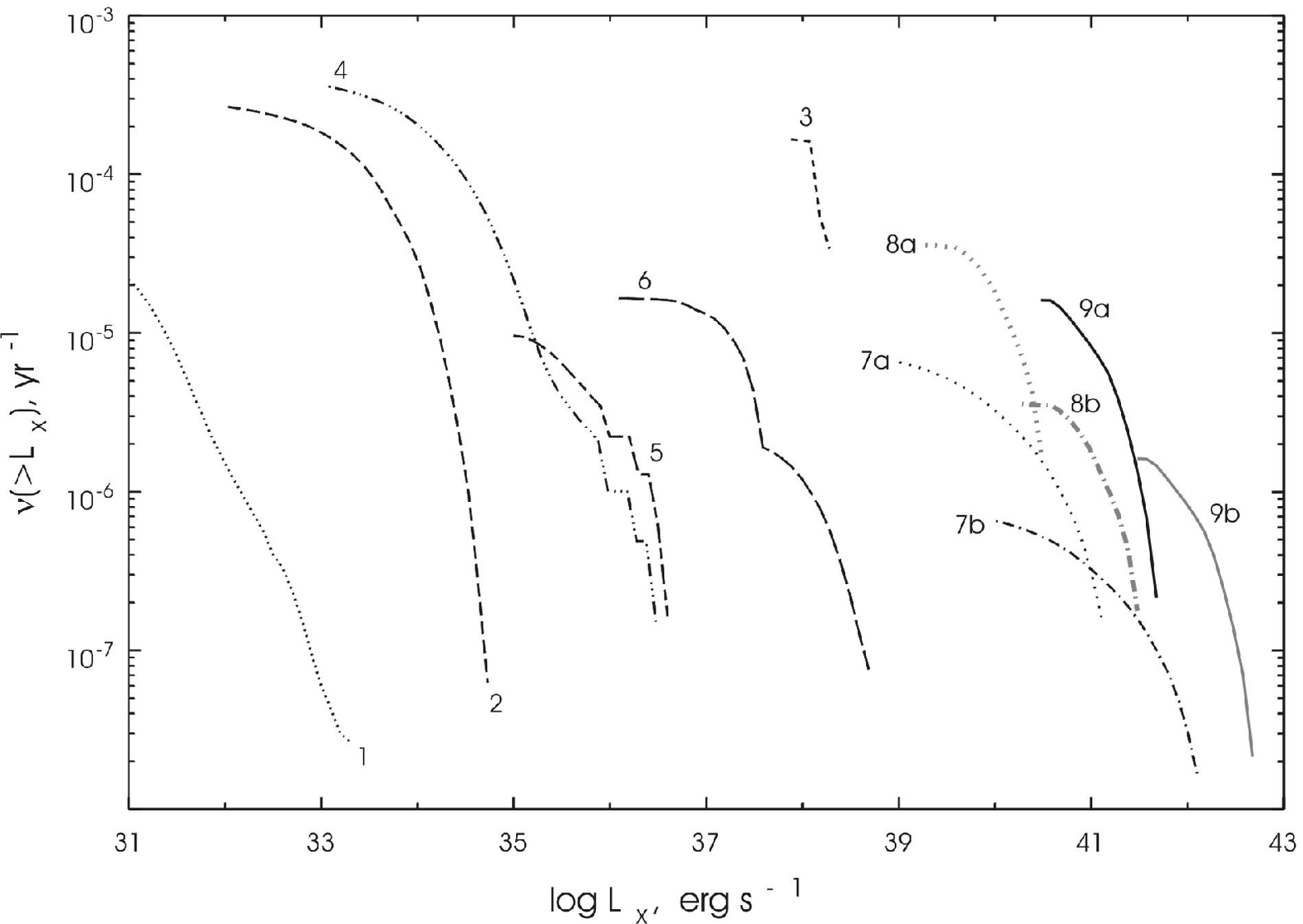} \figcaption{
Birth frequency for different types of X-ray sources in the
Galaxy. Marks in the Figure are: 1, NA+I; 2, NA+II; 3, NA+III; 4,
NA+Be; 5, BH+II; 6, BH+III; 7a, SNA+III, collimation angle
$\alpha=10^{\circ}$; 7b, SNA+III, $\alpha=1^{\circ}$; 8a, SBH+III,
$\alpha=10^{\circ}$; 8b, SBH+III, $\alpha=1^{\circ}$; 9a, SBH+III,
$\alpha=10^{\circ}$; 9b, $\alpha=1^{\circ}$. For curves 9a, 9b
minimal initial mass of the primary star is $120M_{\odot}$, in
other cases it is equal to $10M_{\odot}$. These calculations were
conducted using stellar wind type A. }\label{bfreq1}
\end{figure*}

\begin{figure*}
\epsfxsize=1.0 \textwidth\centering\epsfbox{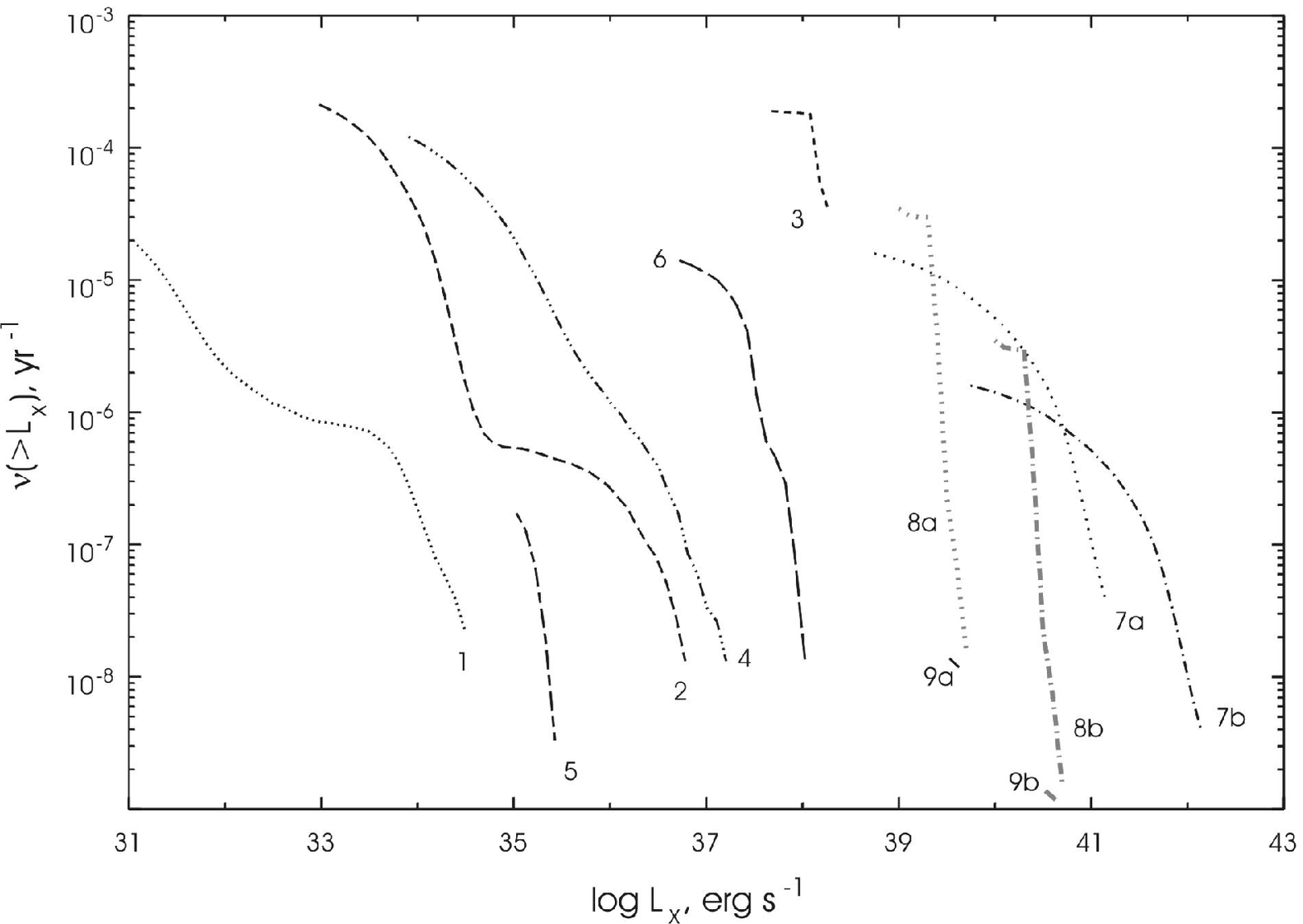} \caption{
Birth frequency for different types of X-ray sources in the
Galaxy. Marks in the Figure are: 1, NA+I; 2, NA+II; 3, NA+III; 4,
NA+Be; 5, BH+II; 6, BH+III, 7a, SNA+III, collimation angle
$\alpha=10^{\circ}$; 7b, SNA+III, $\alpha=1^{\circ}$; 8a, SBH+III,
$\alpha=10^{\circ}$; 8b, SBH+III, $\alpha=1^{\circ}$; 9a, SBH+III,
$\alpha=10^{\circ}$; 9b, $\alpha=1^{\circ}$. For curves 9a, 9b
minimal initial mass of the primary star is $120M_{\odot}$, in
other cases it is equal to $10M_{\odot}$. These calculations were
conducted using stellar wind type C.}\label{bfreq2}
\end{figure*}

\begin{figure*}
\epsfxsize=1.0 \textwidth\centering\epsfbox{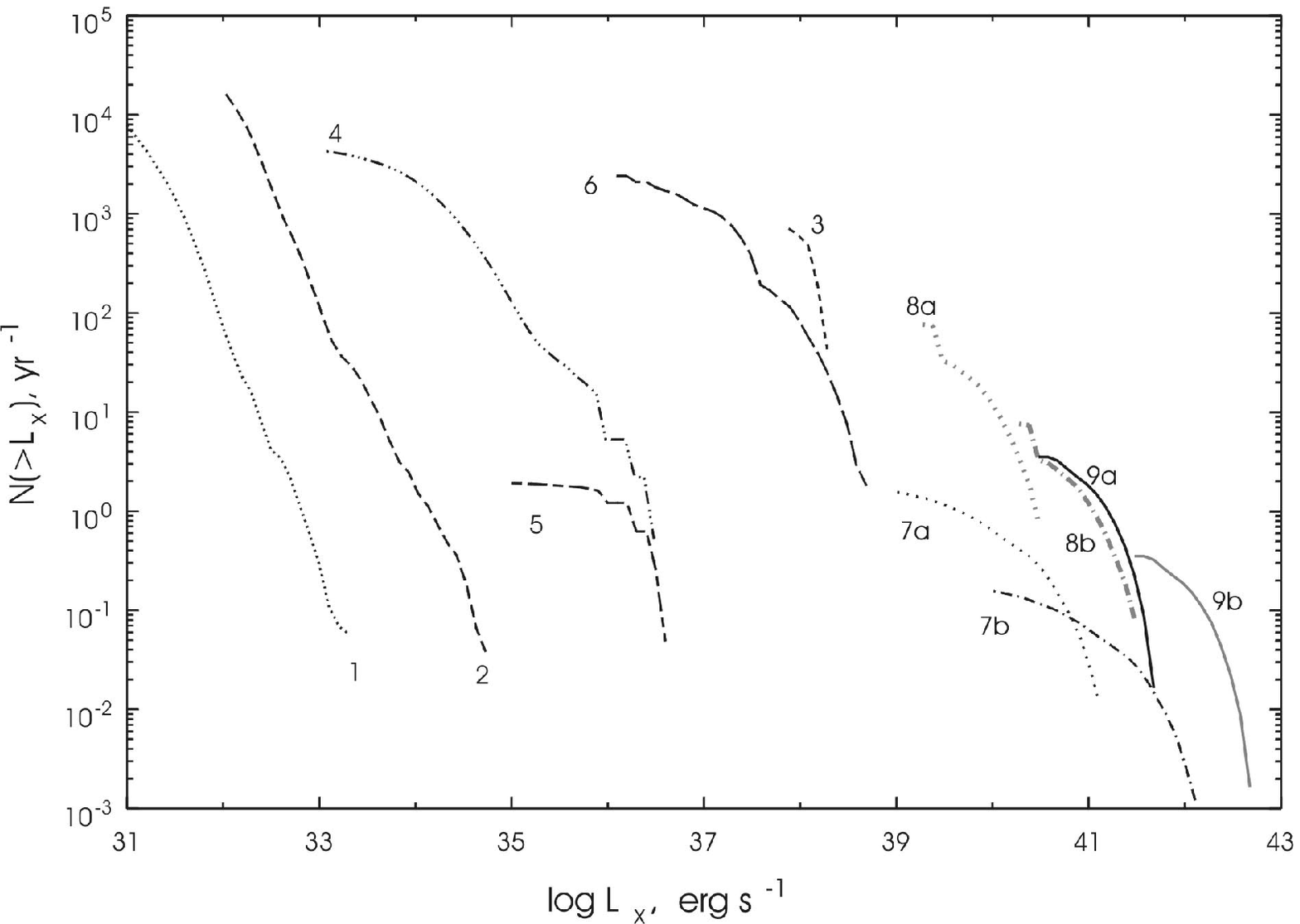} \figcaption{
Cumulative luminosity functions of different types of X-ray
sources in the Galaxy. Marks in the Figure are: 1, NA+I; 2, NA+II;
3, NA+III; 4, NA+Be; 5, BH+II; 6, BH+III; 7a, SNA+III, collimation
angle $\alpha=10^{\circ}$; 7b, SNA+III, $\alpha=1^{\circ}$; 8a,
SBH+III, $\alpha=10^{\circ}$; 8b, SBH+III, $\alpha=1^{\circ}$; 9a,
SBH+III, $\alpha=10^{\circ}$; 9b, $\alpha=1^{\circ}$. For curves
9a, 9b minimal initial mass of the primary star is $120M_{\odot}$,
in other cases it is equal to $10M_{\odot}$. These calculations
were conducted using stellar wind type A. }\label{xlf1}
\end{figure*}

\begin{figure*}
\epsfxsize=1.0 \textwidth\centering\epsfbox{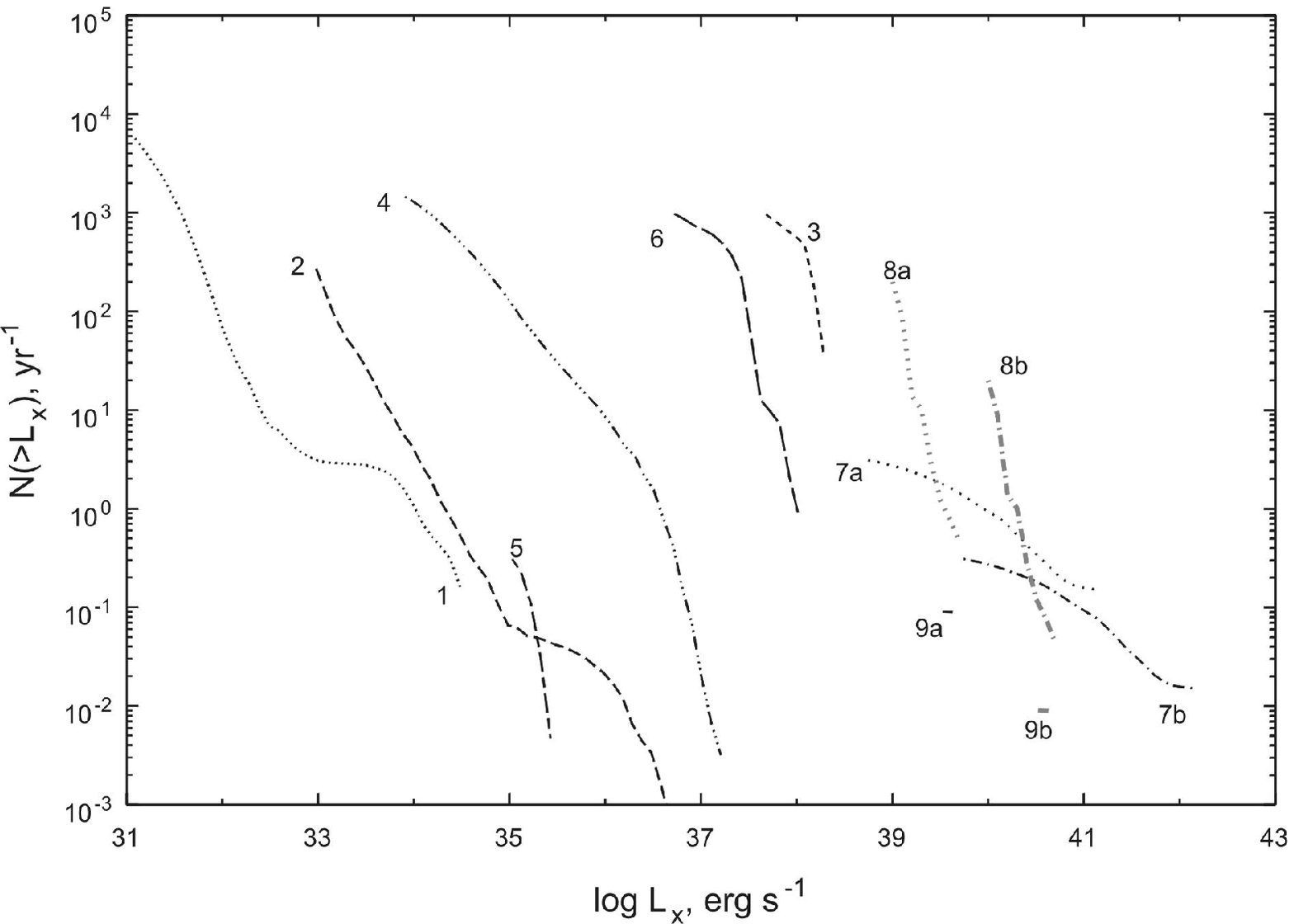} \caption{
Cumulative luminosity functions of different types of X-ray
sources in the Galaxy. Marks in the Figure are: 1, NA+I; 2, NA+II;
3, NA+III; 4, NA+Be; 5, BH+II; 6, BH+III, 7a, SNA+III, collimation
angle $\alpha=10^{\circ}$; 7b, SNA+III, $\alpha=1^{\circ}$; 8a,
SBH+III, $\alpha=10^{\circ}$; 8b, SBH+III, $\alpha=1^{\circ}$; 9a,
SBH+III, $\alpha=10^{\circ}$; 9b, $\alpha=1^{\circ}$. For curves
9a, 9b minimal initial mass of the primary star is $120M_{\odot}$,
in other cases it is equal to $10M_{\odot}$. These calculations
were conducted using stellar wind type C. }\label{xlf2}
\end{figure*}

In the Figures \ref{bfreq1} and \ref{bfreq2} we show birth
frequency of different types of X-ray sources in the spiral
galaxy. In the Figures \ref{xlf1} and \ref{xlf2} we present
cumulative luminosity functions of different types of X-ray
sources in the same galaxy. Figures \ref{bfreq1} and \ref{xlf1}
were calculated using stellar wind type A (weak stellar wind).
Figures \ref{bfreq2} and \ref{xlf2} were calculated using stellar
wind type C (moderate stellar wind). Marks in Figures 4 -- 7 are
(see abbreviation in Section \ref{sec2-sub1}) : 1, NA+I; 2, NA+II;
3, NA+III; 4, NA+Be; 5, BH+II; 6, BH+III; 7a, SNA+III, collimation
angle (for super critical regimes of accretion)
$\alpha=10^{\circ}$; 7b, SNA+III, $\alpha=1^{\circ}$; 8a, SBH+III,
$\alpha=10^{\circ}$; 8b, SBH+III, $\alpha=1^{\circ}$; 9a, SBH+III,
$\alpha=10^{\circ}$; 9b, $\alpha=1^{\circ}$. For curves 9a, 9b
minimal initial mass of the primary star is $120M_{\odot}$, in
other cases it is equal to $10M_{\odot}$.

As one can see from Figures \ref{bfreq1} -- \ref{xlf2}, different
types of X-ray binary systems belong to different luminosity
ranges, their luminosity functions have different slope. These
facts are evidence of complexity of the X-ray luminosity function.

Comparisons between figures 4 and 5, 6 and 7 convince us of the
importance of taking into account life times of X-ray stages in
theoretical models of XLFs. Relative abundances of different types
of X-ray binary systems in the birth frequency function and in the
luminosity function are different. For example, we can see from
Figure 4 that the birth frequency of NA+II X-ray binaries is about
ten times higher than the birth frequency of NA+I X-ray binaries.
But the super giant life time is much shorter than the life time
of the main sequence star, so, as we can see from Figure 5,
quantity of NA+I binaries is only two times less than quantity of
NA+II systems in the spiral galaxy. Stronger stellar wind (type C)
makes this difference even greater (compare Figures 6 and 7).

The stellar wind magnitude essentially influences the scenario for
two reasons. First, the spherically symmetric wind leads to
increase in component separation. Secondly, stellar wind greatly
affects the final evolutionary outcome of massive stars. In
particular, the choice of wind strength will change the mass
distribution of black holes seen in the population
\citep{bogomazov2005a}, as the black hole progenitor loses a
different amount of mass prior to collapse. Moreover, the total
mass loss of a star by wind may cause a change in its remnant type
(it may produce a neutron star instead of a black hole). We can
see from Figures 4 -- 7 that stronger stellar wind (type C)
dramatically decreases quantities of many types of X-ray binaries
(and affects all kind of them).

\begin{figure*}
\epsfxsize=1.0 \textwidth\centering\epsfbox{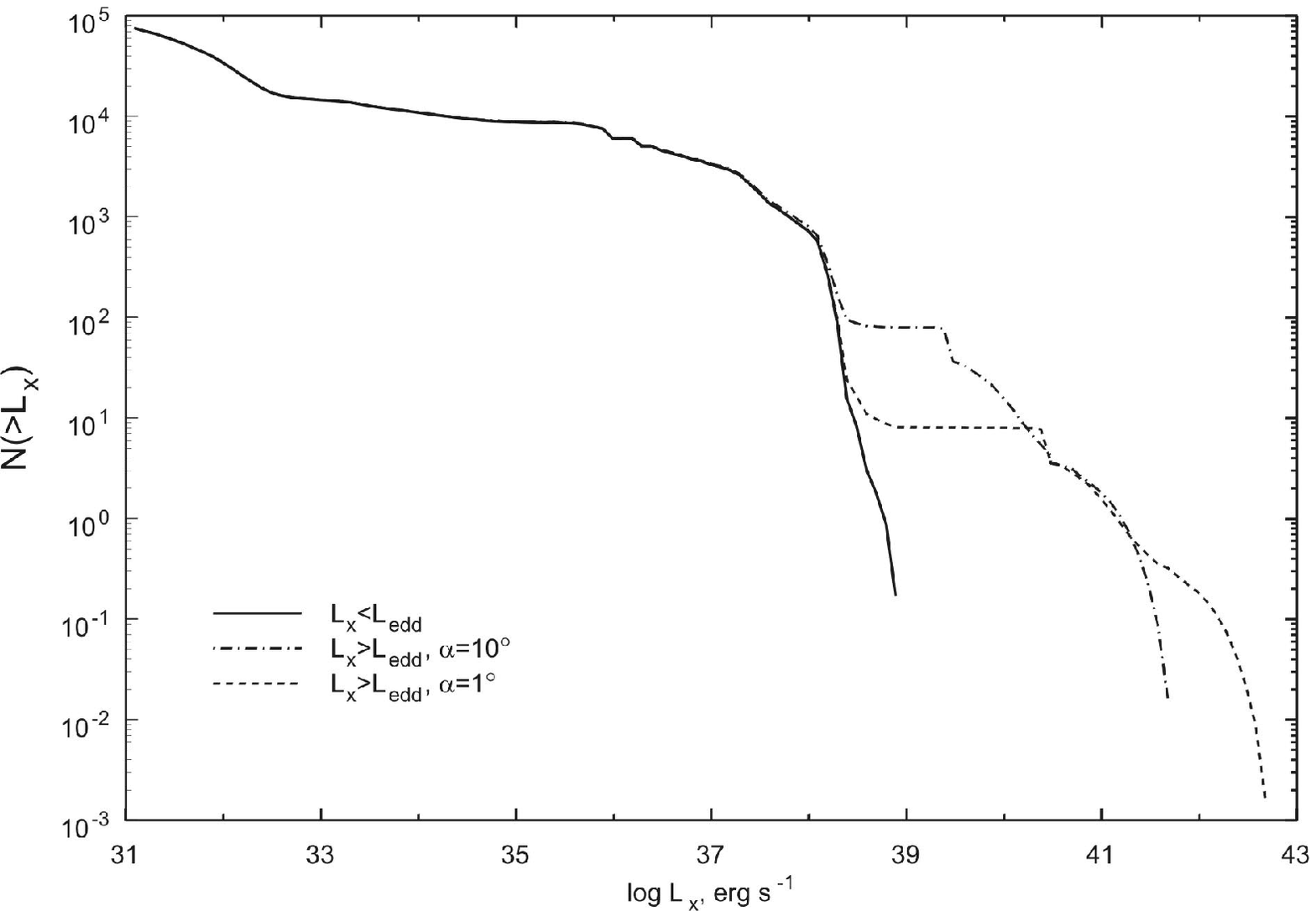} \caption{
Cumulative luminosity functions of all investigated systems in the
galaxy like the Milky Way. See Table 1 for numerical data. In this
Figure $\alpha$ is the collimation angle in supercritical regimes
of accretion. These calculations were conducted using stellar wind
type A. }\label{galxlf1}
\end{figure*}

\begin{figure*}
\epsfxsize=1.0 \textwidth\centering\epsfbox{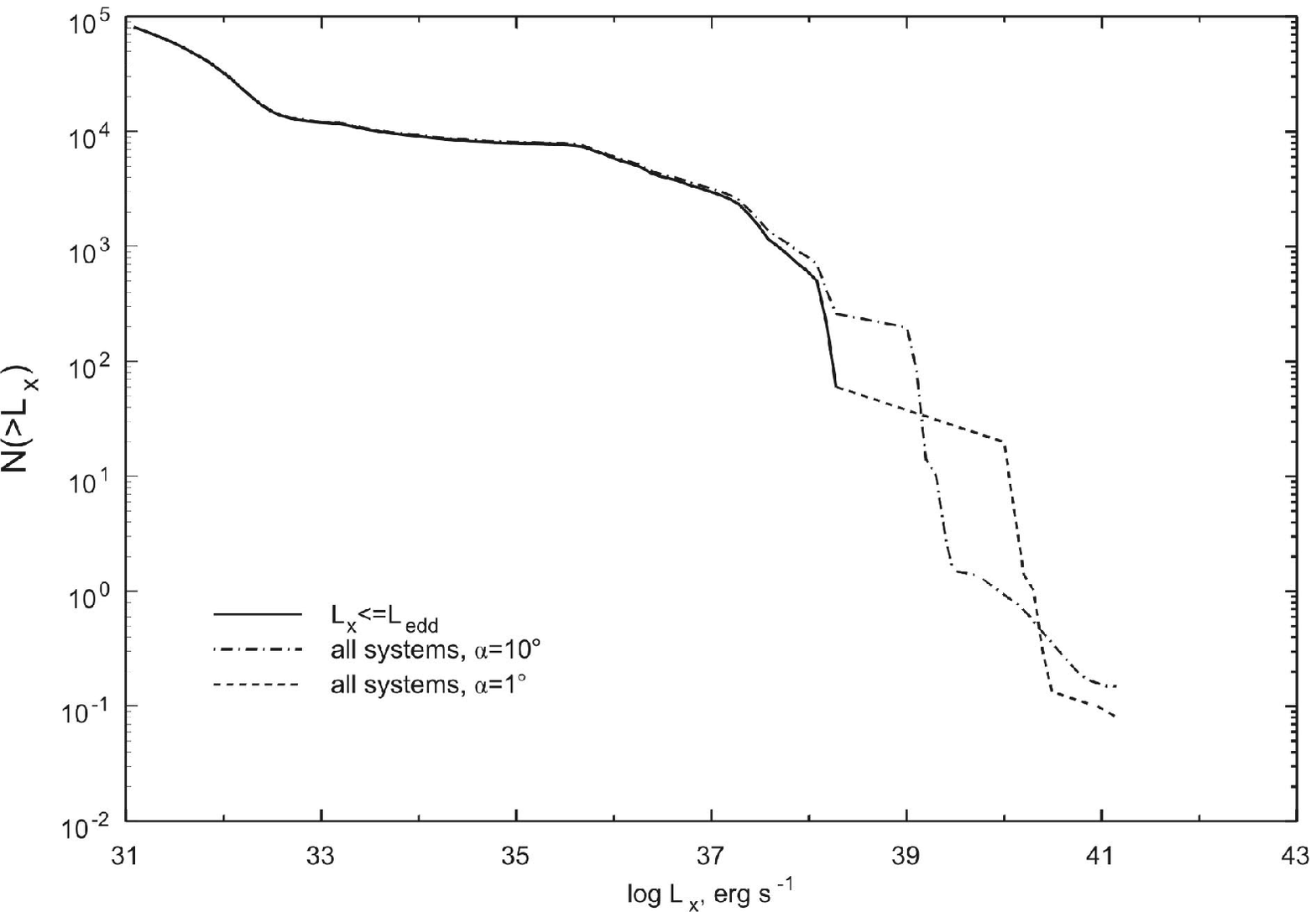} \caption{
Cumulative luminosity functions of all investigated systems in the
galaxy like the Milky Way. See Table 1 for numerical data. In this
Figure $\alpha$ is the collimation angle in supercritical regimes
of accretion. These calculations were conducted using stellar wind
type C. }\label{galxlf2}
\end{figure*}

In the Figures \ref{galxlf1} and \ref{galxlf2} we show cumulative
luminosity functions of all investigated systems in the spiral
galaxy like the Milky Way. See Tables 1 and 2 for numerical data.
In these Figures $\alpha$ is the collimation angle in
supercritical regimes of accretion. Figure \ref{galxlf1} was
calculated under the assumption of stellar wind type A, Figure
\ref{galxlf2} was calculated under the assumption of stellar wind
type C.

Figures \ref{galxlf1} and \ref{galxlf2} show that the X-ray
luminosity function has different slope in different ranges of
luminosity (see also Tables 1 and 2 for numerical data).
\citet{grimm2002a} argued that the best values of the slope and
normalization of the cumulative form of the luminosity function is

\begin{equation}
\label{cumform} N(>L)=5.4 \times SFR \left( L^{-0.61\pm
0.12}-210^{-0.61\pm 0.12}\right);
\end{equation}

\noindent over the luminosity range between $\sim 10^{35}$ erg
s$^{-1}$ and $\sim 10^{40}$ erg s$^{-1}$ (see Figure 5 and
Equation 7 in their paper, but they gave narrower luminosity range
as the result in Conclusions of the article). Our calculations
show similar XLF slope over the ranges between $\approx 2\cdot
10^{37}$ erg s$^{-1}$ and $\approx 10^{38}$ erg s$^{-1}$, and
between $\approx 2\cdot 10^{39}$ erg s$^{-1}$ and $\approx
10^{41}$ erg s$^{-1}$ (the last range depends on our assumptions
about the collimation angle of the X-ray emission in cases of
super critical accretion). Between these two ranges the XLFs (our
theoretical curves) become very steep due to Eddington limit
(there are a lot of NA+III systems, and their luminosity is about
this value, see Figures \ref{bfreq1} -- \ref{xlf2}).

\begin{figure*}
\epsfxsize=0.75 \textwidth\centering\epsfbox{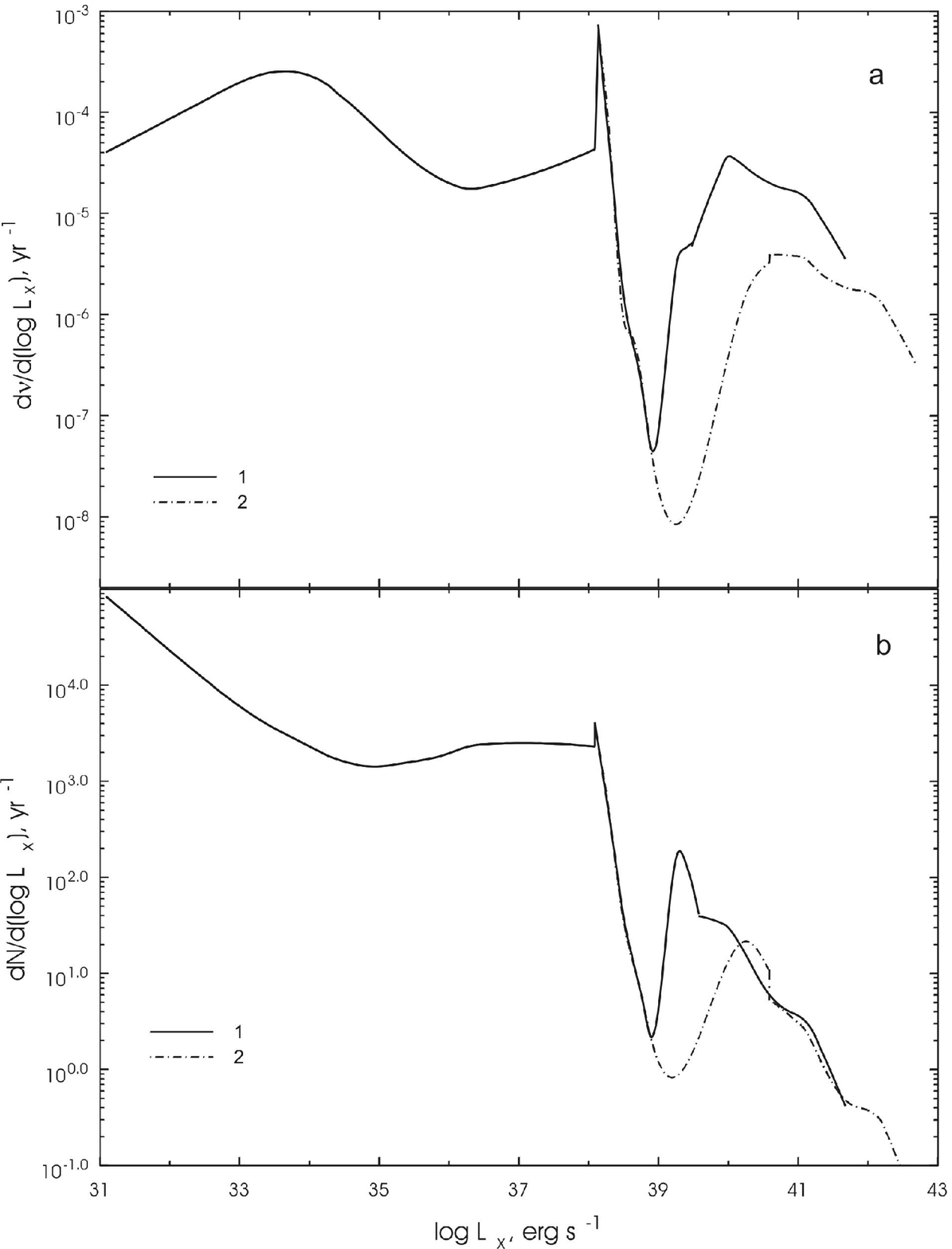} \caption{
Birth frequency (a) of all investigated systems (differential
function) and differential luminosity function (b) of X-ray binary
sources in the Galaxy. Stellar wind type A. Marks in the Figure
are: 1, collimation angle (for super critical regimes)
$\alpha=10^{\circ}$; 2, $\alpha=1^{\circ}$.}\label{dif1}
\end{figure*}

\begin{figure*}
\epsfxsize=0.75 \textwidth\centering\epsfbox{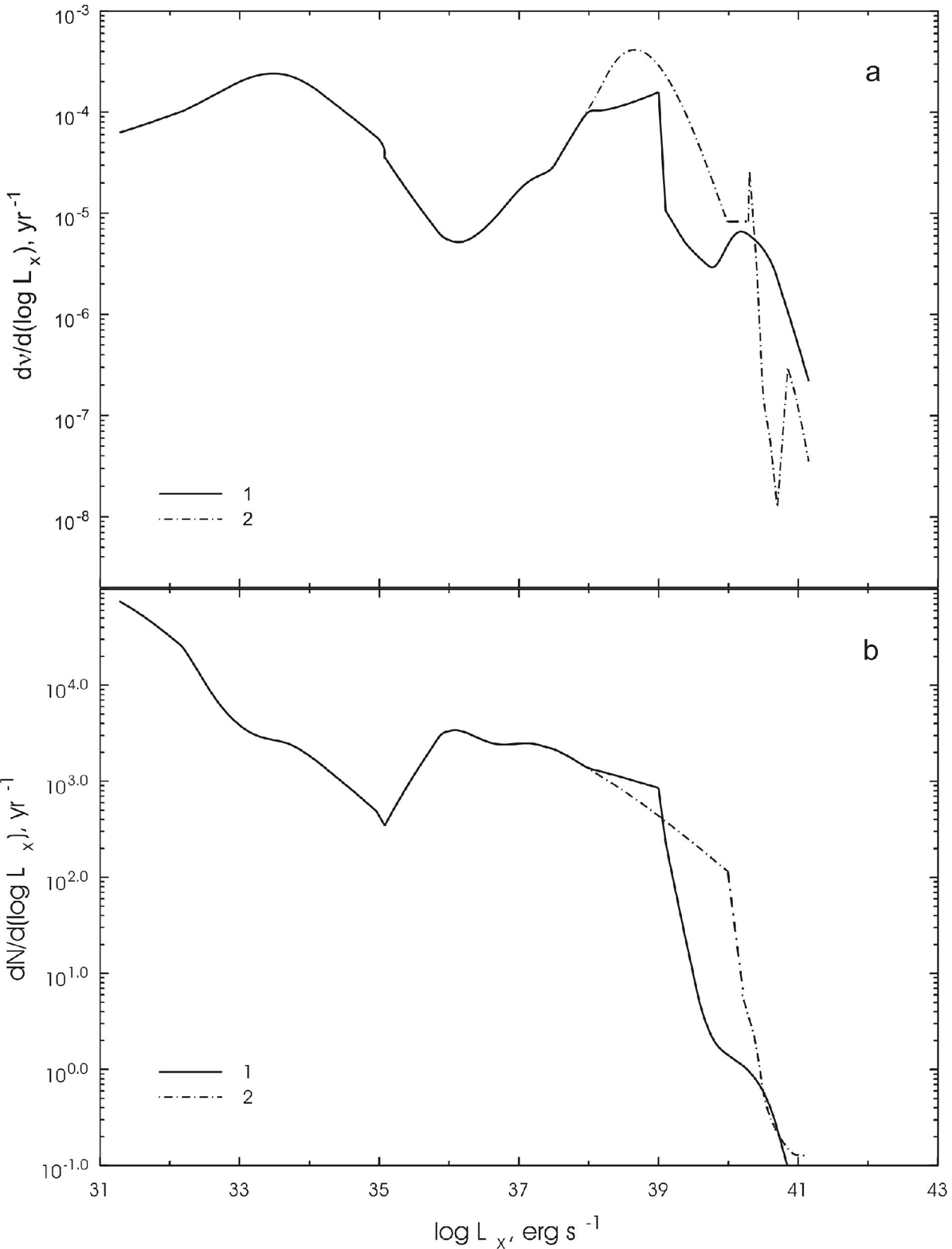} \caption{
Birth frequency (a) of all investigated systems (differential
function) and differential luminosity function (b) of X-ray binary
sources in the Galaxy. Stellar wind type C. Marks in the Figure
are: 1, collimation angle (for super critical regimes)
$\alpha=10^{\circ}$; 2, $\alpha=1^{\circ}$.}\label{dif2}
\end{figure*}

In the Figures \ref{dif1} and \ref{dif2} we show birth frequency
(a) of all investigated systems (differential function) and
differential luminosity function (b) of X-ray binary sources in
the Galaxy. Marks in the Figures are: 1, collimation angle (for
super critical regimes) $\alpha=10^{\circ}$; 2,
$\alpha=1^{\circ}$. Figure \ref{dif1} was calculated under the
assumption of stellar wind type A, Figure \ref{dif2} was
calculated under the assumption of stellar wind type C. Luminosity
functions in differential form also have different slope, there is
no evidence for the universal XLF.

\begin{figure*}
\epsfxsize=0.35 \textwidth\centering\epsfbox{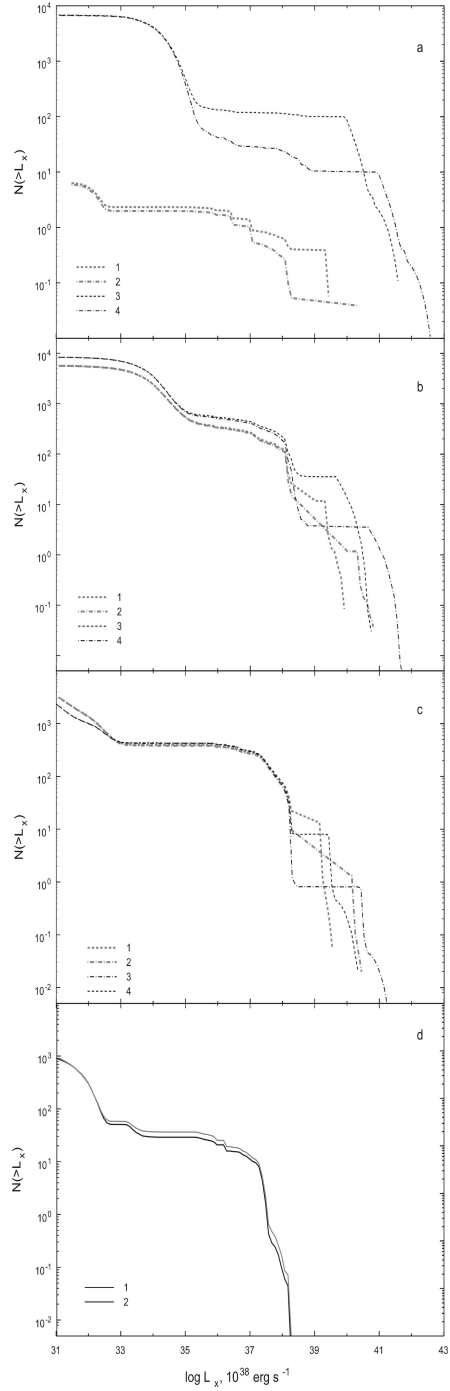} \caption{
Cumulative luminosity functions of all investigated systems in the
``elliptical'' galaxy after the star formation burst. See Table 1
for numerical estimations. The curves in the Figure represent the
next models: 1, stellar wind type A, collimation angle (for super
critical regimes) $\alpha=10^{\circ}$; 2, wind A,
$\alpha=1^{\circ}$; 3, wind C, $\alpha=10^{\circ}$; 4, wind C,
$\alpha=1^{\circ}$. The time ranges after the star formation burst
in the Figure are: a, 0-10 million years; b, 10-100 million years;
c, 100 million -- 1 billion years; d, 1-10 billion
years.}\label{ellip}
\end{figure*}

In the Figure \ref{ellip} we show cumulative luminosity functions
of all investigated systems in the elliptical galaxy after the
star formation burst. The curves in the Figure represent the next
models: 1, stellar wind type A, collimation angle (for super
critical regimes) $\alpha=10^{\circ}$; 2, wind A,
$\alpha=1^{\circ}$; 3, wind C, $\alpha=10^{\circ}$; 4, wind C,
$\alpha=1^{\circ}$. The time ranges after the star formation burst
in the Figure are: a, 0-10 million years; b, 10-100 million years;
c, 100 million -- 1 billion years; d, 1-10 billion years. v Figure
\ref{ellip} shows the evolution of luminosity function on a long
timescale after the stellar formation burst in the elliptical
\footnote{In this work we treat the galaxy as ``elliptical'' if
the object has mass $10^{11}M_{\odot}$ and $\delta$-function
starburst.} galaxy. As one can see from this Figure, there is no
evidence for the universal XLF. Nevertheless, note that numbers of
systems in this figure are quite relative. Any systems were added
to the number of the systems in appropriate interval of time if
they born as X-ray system or still show itself as X-ray source
during this period or part of it, but life time of a system can be
less than the duration of the period and a system can born not in
the beginning of the period. For more precision it is necessary to
take less intervals of time, but our purpose is to show long time
evolution qualitatively.

\citet{belc2004} found that the dependence of the XLF slope on age
is non-monotonic in the dwarf (post)starburst galaxy NGC 1569.
They studied behavior with time of theoretical normalized XLFs for
two stellar populations: one old at 1.5 Gyr and one young at age
10, 70, and 170 Myr (continuous SFR through 1.5 Gyr, and 10, 70,
and 100 Myr, respectively). The average SFR in the old population
was assumed to be 20 times smaller than that in the young
population. Direct comparison between our results is difficult,
because we use different star formation models in our
calculations. One of the common features is in that that the XLF
should evolve with time. Also we suggest that their XLFs can be
fitted by broken power laws, \citet{belc2004} did not obtain
uniform XLF in NGC 1569.

\begin{figure*}
\epsfxsize=1.0 \textwidth\centering\epsfbox{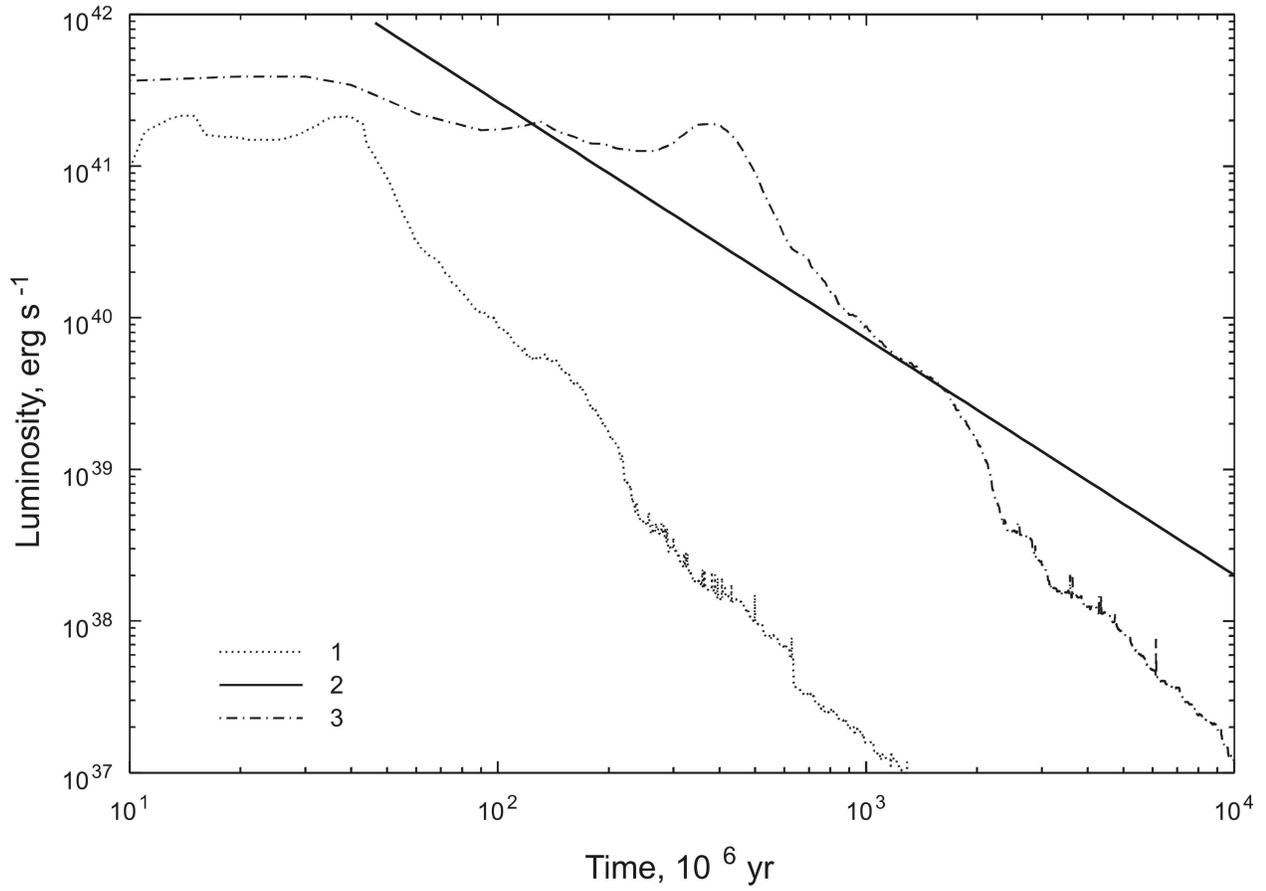} \caption{
Evolution of the X-ray luminosity after the star formation burst
($T=0$) in the galaxy with mass $10^{11} M_{\odot}$. See Table 2
for numerical data. In this Figure: 1, our calculations, stellar
wind type A; 2, the result obtained by \citet{tatarintzeva1989a};
3, our calculations, stellar wind type C. }\label{lum}
\end{figure*}

In the Figure \ref{lum} we show the evolution of the X-ray
luminosity after the star formation burst ($T=0$) in the galaxy
with mass $10^{11} M_{\odot}$. See Table 2 for numerical data. In
this Figure: 1, our calculations, stellar wind type A; 2, the
result obtained by \citet{tatarintzeva1989a}; 3, our calculations,
stellar wind type C. We should note that in comparison with
results obtained by \citet{vanbever2000} we do not take into
account the X-ray emission from supernova remnants in our models.
Our data in this Figure start at their end point (10 Myr). After
$4\cdot 10^2$ million years since star formation burst in the
galaxy its X-ray luminosity can be rather well fitted by power law
$L(T)\sim T^{-a}$; $a$ is equal to $1.56$ and $1.8$ in very wide
range of time (see Table 3 for details). Previous work
\citep{tatarintzeva1989a} showed approximately the same result
which we can confirm. The cause of differences is in that that 16
years ago calculations were conducted if authors were taking into
consideration not so much types of systems as in the present work.
Also models of evolution of binaries have changed. Stronger
stellar wind (see Table 4) makes the our result almost
inconsistent with \citet{tatarintzeva1989a}.

So, our calculations show the next results:

\begin{enumerate}

\item X-ray luminosity function of binary X-ray sources is
complicated, it has different slope in different ranges of
luminosity. So, there is no universal X-ray luminosity function of
binary X-ray sources.

\item X-ray luminosity function of binary X-ray sources depends on
the star formation rate as it was first shown in 1989
\citep{tatarintzeva1989a}.

\item It is necessarily to take into account spin evolution of
neutron stars and life times of all stages during theoretical
modelling of X-ray luminosity function of binary X-ray sources.

\end{enumerate}

\newpage

\newpage

\begin{table*}
\begin{center} \caption{Numerical approximation of the cumulative
luminosity function in the spiral galaxy. Stellar wind type A. See
Figure \ref{galxlf1} for graphical data. }
\begin{tabular}{ll}
\tableline \tableline
 Luminosity range, & $k$\tablenotemark{a} \\
 $\log L_x$, erg s$^{-1}$ & \\
\tableline
$31.0$ -- $32.5$ & $-0.25$ \\
$32.5$ -- $35.6$ & $-0.1$ \\
$35.6$ -- $37.2$ & $-0.25$ \\
$37.2$ -- $38.0$ & $-0.7$ \\
$38.0$ -- $38.3$ & $\approx -8$\tablenotemark{b} \\
$38.0$ -- $38.5$ & $\approx -8$\tablenotemark{c} \\
$38.3$ -- $39.2$ & $\approx -0.05$\tablenotemark{b} \\
$38.5$ -- $40.2$ & $\approx -0.05$\tablenotemark{c} \\
$39.2$ -- $41.1$ & $-0.7$\tablenotemark{b} \\
$40.2$ -- $42.2$ & $-0.7$\tablenotemark{c} \\
\tableline
\end{tabular}
\tablenotetext{a}{fit curve is $N(>L)\sim L^{k}$.}
\tablenotetext{b}{collimation angle (for super critical regimes)
$\alpha=10^{\circ}$.} \tablenotetext{c}{collimation angle (for
super critical regimes) $\alpha=1^{\circ}$.}
\end{center}
\end{table*}

\begin{table*}
\begin{center} \caption{Numerical approximation of the cumulative
luminosity function in the spiral galaxy. Stellar wind type C. See
Figure \ref{galxlf2} for graphical data. }
\begin{tabular}{ll}
\tableline \tableline
 Luminosity range, & $k$\tablenotemark{a} \\
 $\log L_x$, erg s$^{-1}$ & \\
\tableline
$31.0$ -- $32.5$ & $-0.4$ \\
$32.5$ -- $35.5$ & $-0.1$ \\
$35.5$ -- $37.2$ & $-0.3$ \\
$37.2$ -- $38.0$ & $-0.6$ \\
$38.0$ -- $38.2$ & $-1.8$\tablenotemark{b} \\
$38.0$ -- $38.2$ & $-3.6$\tablenotemark{c} \\
$38.2$ -- $39.1$ & $-0.1$\tablenotemark{b} \\
$38.2$ -- $40.1$ & $-0.3$\tablenotemark{c} \\
$39.1$ -- $39.5$ & $-3.5$\tablenotemark{b} \\
$40.1$ -- $40.5$ & $-3.5$\tablenotemark{c} \\
$39.5$ -- $41.0$ & $-0.75$\tablenotemark{b} \\
$40.5$ -- $42.0$ & $-0.75$\tablenotemark{c} \\
\tableline
\end{tabular}
\tablenotetext{a}{fit curve is $N(>L)\sim L^{k}$.}
\tablenotetext{b}{collimation angle (for super critical regimes)
$\alpha=10^{\circ}$.} \tablenotetext{c}{collimation angle (for
super critical regimes) $\alpha=1^{\circ}$.}
\end{center}
\end{table*}

\begin{table*}
\begin{center} \centering
\caption{Numerical approximation of the X-ray luminosity of the
galaxy after the star formation burst. Stellar wind type A. }
\begin{tabular}{lll}
\tableline
Time range, & $c_1$\tablenotemark{a} & $p$\tablenotemark{a} \\
$10^6$ yr & & \\
\tableline
$4\cdot 10^2$ -- $1\cdot 10^3$ & $3\cdot 10^{47}$ & $-2.5$ \\
$1\cdot 10^3$ -- $2\cdot 10^3$ & $3.6\cdot 10^{44}$ & $-1.56$ \\
$2\cdot 10^3$ -- $2.5\cdot 10^3$ & $2\cdot 10^{52}$ & $-4$ \\
$2.5\cdot 10^3$ -- $1\cdot 10^4$ & $3\cdot 10^{44}$ & $-1.8$ \\
\tableline
\end{tabular}
\tablenotetext{a}{fit curve is $L(T)=c_1 (T/10^6 \mbox{yr})^{p}$
erg s$^{-1}$.}
\end{center}
\end{table*}

\begin{table*}
\begin{center} \centering
\caption{Numerical approximation of the X-ray luminosity of the
galaxy after the star formation burst. Stellar wind type C. }
\begin{tabular}{lll}
\tableline
Time range, & $c_1$\tablenotemark{a} & $p$\tablenotemark{a} \\
$10^6$ yr & & \\
\tableline
$4\cdot 10$ -- $1\cdot 40$ & $1.5\cdot 10^{41}$ & $\approx 0$ \\
$1\cdot 40$ -- $1.5\cdot 10^3$ & $2\cdot 10^{45}$ & $-2.7$ \\
\tableline
\end{tabular}
\tablenotetext{a}{fit curve is $L(T)=c_1 (T/10^6 \mbox{yr})^{p}$
erg s$^{-1}$.}
\end{center}
\end{table*}


\begin{thebibliography}{99}

\bibitem[Abt (1983)]{abt1983a} Abt, H. A. 1983, ARA\&A, 21, 343

\bibitem[ATNF
catalogue (2006)]{atnf} Manchester, R. N., Hobbs, G. B., Teoh, A.
\& Hobbs, M. 2005 (1993-2006), AJ, 129 (the Australia Telescope
National Facility (ATNF) pulsar catalogue, avialable at:
http://www.atnf.csiro.au/research/pulsar/psrcat/)

\bibitem[Belczynski et al. (2004)]{belc2004} Belczynski, K., Kalogera, V.,
Zezas, A., \& Fabbiano, G. 2004, ApJ, 601, 147


\bibitem[Bogomazov et al. (2005)]{bogomazov2005a} Bogomazov, A. I., Abubekerov M. K., Lipunov, V.
M.2005, Astronomy Reports, 49, 644

\bibitem[Dewi \& Tauris (2000)]{dewi2000} Dewi, J. D. M., \& Tauris, T. M. 2000, A\&A,
360, 1043

\bibitem[Georgantopoulos et al. (2005)]{georgantopoulos2005a} Georgantopoulos, I.,
Georgakakis, A., \& Koulouridis, E. 2005, MNRAS, 360, 782

\bibitem[Gilfanov (2004)]{gilfanov2004a} Gilfanov, M. 2004, MNRAS, 349, 146

\bibitem[Grimm et al. (2002)]{grimm2002a} Grimm, H.-J., Gilfanov, M., \& Sunyaev, R. 2002, A\&A, 391, 923

\bibitem[Grimm et al. (2003)]{grimm2003a} Grimm, H.-J., Gilfanov, M., \& Sunyaev, R. 2003, MNRAS, 339, 793

\bibitem[Hobbs et al. (2005)]{hobbs2005} Hobbs, G., Lorimer, D. R., Lyne, A. G., \& Kramer, M. 2005, MNRAS, 360,
974

\bibitem[Karpov \& Lipunov (2001)]{karpov} Karpov, S. V., \& Lipunov, V. M. 2001, Astron. Letters, 2001, 27, 10, 645-647

\bibitem[Kim \& Fabbiano (2004)]{kim2004a} Kim, D.-W., \&
Fabbiano, G. 2004, ApJ, 611, 846

\bibitem[King et al. (2001)]{king2001} King, A. R., Davies, M. B., Ward, M. J., Fabbiano,
G. \& Elvis, M. 2001, ApJ, 552, L109

\bibitem[Kong (2003)]{kong2003a} Kong,
A. K. H. 2003, MNRAS, 346, 265

\bibitem[Lipunov (1982a)]{lipunov1982a} Lipunov, V. M. 1982a, Ap\&SS,
82, 343

\bibitem[Lipunov (1982b)]{lipunov1982b} Lipunov, V. M. 1982b, SvA, 26, 54

\bibitem[Lipunov (1982c)]{lipunov1982c} Lipunov, V. M. 1982c, SvAL, 8, 194

\bibitem[Lipunov (1992)]{lipunov1992a} Lipunov, V. M. 1992, Astrophysics of Neutron Stars, Springer-Verlag, Berlin -
Heidelberg - New York, Astronomy and Astrophysics Library, 322

\bibitem[Lipunov et al. (1996a)]{lipunov1996a} Lipunov, V. M., Ozernoy, L. M., Popov, S. B., Postnov, K. A.,
\& Prokhorov, M. E. 1996a, ApJ, 466, 234

\bibitem[Lipunov et al. (1996b)]{lipunov1996b} Lipunov, V. M., Postnov, K. A., \& Prokhorov, M. E. 1996b, ed. R. A. Sunyaev, The Scenario Machine: Binary Star Population
Synthesis, Astrophysics and Space Physics Reviews, vol. 9, Harwood
academic publishers

\bibitem[Lipunov et al. (1996c)]{lipunov1996c} Lipunov, V. M., Postnov, K. A., \& Prokhorov, M. E.
1996, A\&A, 310, 489

\bibitem[Lipunov et al. (1997)]{lipunov1997a} Lipunov, V. M., Postnov, K. A., \& Prokhorov, M. E., 1997, MNRAS, 288, 245

\bibitem[Lipunov (2006)]{lipunov2006a} Lipunov, V. M. 2006, IAU proseedings,
Populations of High Energy Sources in Galaxies Proceedings of the
230th Symposium of the International Astronomical Union, Edited by
E. J. A. Meurs, G. Fabbiano, Cambridge University Press, 2006, p.
391

\bibitem[Lipunov et al. (2007)]{lipunov2007a} Lipunov, V. M.,
Postnov, K. A., Prokhorov, M. E., Bogomazov A. I. 2007,
arXiv:0704.1387v1

\bibitem[Liu et al. (2000)]{liu2000a} Liu, Q. Z., van Paradijs, J., \& van den
Heuvel, E. P. J. 2000, A\&A, 147, 25

\bibitem[Lij et al. (2001)]{liu2001a} Liu, Q. Z., van Paradijs, J., \& van den
Heuvel, E. P. J. 2000, A\&A, 368, 1021

\bibitem[Muno et al. (2004)]{muno2004a} Muno, M. P. 2004, ApJ, 613, 1179

\bibitem[Popov et al. (1998)]{popov1998a} Popov, S. B., Lipunov, V. M., Prokhorov, M. E., \& Postnov, K. A. 1998, Astronomy Reports,
v. 42, p. 29

\bibitem[Postnov (2003)]{postnov2003a} Postnov, K. A. 2003, Astron. Lett., 29, 372

\bibitem[Raguzova \& Popov (2005)]{raguzova2005a} Raguzova, N. V., \& Popov
S. B. 2005 Astronomical and Astrophysical Transactions, 24, 151

\bibitem[Rappaport et al. (2005)]{rappaport2005a} Rappaport, S. A., Podsiadlowski, Ph., \& Pfahl, E. 2005, MNRAS, 356, 401

\bibitem[Tatarinzeva et al. (1989)]{tatarintzeva1989a} Tatarintzeva, V., Lipunov, V., Osminkin, E., \& Prokhorov
M. E. 1989, In ESA, The 23rd ESLAB Symposium on Two Topics in
X-Ray Astronomy, v.1: X Ray Binaries, p. 653

\bibitem[Van Bever \& Vanbeveren (2000)]{vanbever2000} Van Bever, J., \& Vanbeveren, D. 2000, A\&A, 358, 462

\bibitem[van den Heuvel (1994)]{heuvel1994a} van den Heuvel, E. P. J. 1994, in Shore S.N., Livio M., van den Heuvel E.P.J., Interacting Binaries, Springer-Verlag, p. 103

\bibitem[Zezas \& Fabbiano (2002)]{zezas2002a} Zezas, A., \& Fabbiano, G. 2002, ApJ, 577, 726

\bibitem[Zezas et al. (2004)]{zezas2004a} Zezas, A., Fabbiano, G., Baldi, A., King, A. R., Ponman, T. J., Raymond, J. C., \& Schweizer, F. 2004, RevMexAA, 20, 53

\bibitem[Cherepashchuk et al. (2005)]{cherepashchuk2005a} Cherepashchuk, A. M., et al. 2005, A\&A, 437, 561

\bibitem[Georgakakis et
al. (2004)]{georgakakis2004a} Georgakakis, A., et al. 2004, MNRAS,
349, 135

\bibitem[Grindlay et al. (2005)]{grindlay2005a} Grindlay, J. E., et al. 2005, ApJ, 635, 920

\end{thebibliography}
\end{document}